\documentclass[12pt,a4paper]{article}
\usepackage{graphicx}
\begin{document}
\begin{center}
\noindent
{\Large \bf On The Role Of The Interface Charge In Non-Ideal Metal-Semiconductor Contacts
}\vspace{5mm}

\bf Dean Koro\v sak and Bruno Cvikl\\[5mm]

{\bf Chair for Applied Physics, Faculty of Civil Engineering\\University of Maribor, Maribor Slovenia\\
and "J. Stefan" Institute, Ljubljana Slovenia\\
email: dean.korosak@uni-mb.si}
\end{center}

\begin{abstract}
The bias dependent interface charge is considered as the origin of the observed non-ideality in current-voltage
and capacitance-voltage characteristics. Using the simplified model for the interface electronic structure based on
defects interacting with the continuum of interface states, the microscopic origin of empirical parameters describing the
bias dependent interface charge function is investigated. The results show that in non-ideal metal-semiconductor 
contacts the interface charge function depends on the interface disorder parameter, density of defects,  
barrier pinning parameter and the effective gap center. The theoretical predictions are tested against several sets of published 
experimental data on bias dependent ideality factor and excess capacitance in various metal-semicoductor systems.
\end{abstract}

\section{Introduction}

The Schottky barrier heights at the metal-semiconductor interfaces
are usually found to be only weakly dependent on the metal work function~\cite{bardeen}. 
The failure to comply with the Schottky-Mott rule~\cite{rhoderick,sze}, where the Schottky barrier height, $\phi_{b}$, 
is equal to the difference between the
metal work function, $\phi_{m}$, and semiconductor electron affinity, $\chi_{s}$, 
is often termed Fermi level pinning.   
This phenomenon has stimulated extensive debate over the past few
decades, resulting in several proposed models, and has recently again attracted 
considerable attention in the scientific community~\cite{tung00,hasegawa,korosak02,korosak01}.  
Interface properties enter these models in the form of the charge localized near the
interface resulting from the finite density of states in the band gap and/or defect levels.
It has been recently shown~\cite{cvikl01,cvikl02,korosak02} 
that delibarate introduction of metal atoms into the thin semiconductor layer near the 
metal-semicoductor interface utilizing ionized cluster beam 
deposition~\cite{takagi,brown,mceachern,cvikl01}, enables identification of the origin and
analysis of non-ideality in certain metal-semiconductor contacts. 

The non-ideality is a consequence of the interlayer present in the structure between the metal and
the semiconductor substrate. In ionized cluster beam deposited 
samples, 
this interlayer consists of
the penetrated metal atoms embedded in the host semiconductor lattice where they introduce 
disorder into the lattice and may act as additional deep impurities. The incident beam in ionized cluster beam 
deposition is comprised of neutral and, to smaller extent, also ionized metal atoms~\cite{takagi,brown,mceachern}.
The small number of neutral and ionized metal clusters of various sizes which are simultaneously present in the incident
beam, on account of the very low energy shared by each cluster atom (of the order of about 60 eV per atom for an
avarage cluster size of 16 atoms and 1 keV translational energy) can be safely neglected. The additionally gained 
energy, up to few keV in some cases, enables the metal ions to enter into semiconductor bulk and come to rest within 
and up to their penetration length ($L$) typically a few nm in depth. 
In such a non-ideal metal-semiconductor structure 
one can identify two distinct interfaces, 
one situated at the metal/disordered interlayer, and the other at the disordered layer/regular semiconductor 
junction~\cite{cvikl02}. 
It was shown that the electronic properties of such a non-ideal Schottky structure determine its low-frequency 
capacitance and Fermi level pinning behaviour~\cite{korosak02,cvikl02}.

Closely connected to the questions of the microscopic properties of the m/s interface is
the problem of the excess capacitance~\cite{werner,ho,wu1,gomila,cvikl02,cvikl01} 
in Schottky junctions. 
Recently~\cite{cvikl02} we have been able to 
show that the observed excess capacitances in non-ideal Schottky junctions arise on account
of the bias dependent localized charge density induced at the appropriate interface in the structure.
The model function for the interface charge 
was chosen to be:
\begin{equation}
\sigma(U)=b\exp [-(K+qU/U_{0})^{2}],
\label{cf}
\end{equation} 
where the constants $b$, $K$, and $U_{0}$ were determined by comparison with the experimental capacitance-voltage 
dependence of different systems~\cite{cvikl02}.
The differential capacitance is in this model given by~\cite{cvikl02}:
\begin{equation}
C(U)=\frac{\epsilon_{s}}{W}+\left(\frac{L}{W}-1\right)\frac{d\sigma}{dU},
\label{cap}
\end{equation}
where $L$ is the width of the disordered layer, and $W$ is the width of the depletion layer. 
Here, the dielectric constants of the disordered layer and the homogenuous 
semiconductor ($\epsilon_{s}$) are taken to be equal.
As an example, we show on figure 1 the measured low frequency capacitance (plotted as $C^{-2}(U)$) of e-beam deposited Al/n-Si
(data taken from refs.~\cite{tseng1,tseng2}) and of ICB deposited Pb/p-Si (our own samples) Schottky structures
as a function of applied bias. The full line is the calculated capacitance-voltage characterstics based on eqs.~\ref{cf} and~\ref{cap}.
For further details we refer the reader to reference~\cite{cvikl02}. 

The current-voltage characteristics of non-ideal metal-semiconductor contacts do not usually follow the characteristics 
predicted by thermionic emission transport theory. The observed discrepance is often treated by introducing the ideality factor
into the current-voltage characteristic defined by~\cite{rhoderick}:
\begin{equation}
n^{-1}=1-\frac{d\phi_{b}}{qdU},
\label{def-if}
\end{equation}
where $\phi_{b}$ is the barrier height.

In what follows we show that the bias dependence of the ideality factor used to describe the non-ideality of the
current-voltage characterstics can be traced back to the interface charge causing the non-ideality and later 
on we discuss the microscopic origin of the parameters in the
interface charge functions. 
The derived expressions are used on 
experimental sets of data published by different authors, and some of the values of microscopic parameters are
extracted.
 
\section{Ideality factor and interface charge in non-ideal Schottky contacts} 

In a non-ideal metal-semiconductor system we can quite generally describe the non-ideality with 
the macroscopic charge situated near the metal-semiconductor interface. The charge neutrality 
of the system demands:
\begin{equation}
Q_{m}+Q_{sc}+Q_{i} =0,
\end{equation}
where $Q_{m}$, $Q_{sc}$, and $Q_{i}$ are the surface charge densities on the metal in the
depletion region of the semiconductor, and at the interface. 

For the non-ideal metal-semiconductor structures with the disordered interlayer and the bias dependent 
charge density the charge neutrality is written as:
\begin{equation}
Q_{m}+Q_{sc}+Q_{i} =Q_{m}+qn_{2}W+q(n_{2}+n_{1})L+\sigma (U)=0,
\end{equation}
where $n_{2}$ is the concentrations of the ionized bulk semiconductor dopants, $n_{1}$ 
is the concentrations of the ionized defects in the disordered interlayer, and $L$ is the width of the
disordered interlayer. The bias dependent localized charge density residing at the
disordered layer/homogenous semiconductor interface is here introduced with $\rho_{i}=\sigma (U)\delta (z-L)$. 

The electrostatic analysis of the model yielded the following expression for 
the width of the depletion layer: 
\begin{equation}
W=\sqrt{L^{2}\left(1-\frac{n_{1}}{n_{2}}\right)
+\frac{\epsilon_{s}}{q^{2}n_{2}}\left(\frac{-q\sigma}{\epsilon_{s}}L-qU+\phi_{b,0}-\chi_{s}\right)},
\end{equation}
where $U$ is the applied bias, $\phi_{b,0}$ is the Schottky barrier height 
(potential barrier at the metal/disordered layer interface).

The potential at the disordered layer/semiconductor interface is defined as 
the effective Schottky barrier height in non-ideal contact:
\begin{equation}
\phi_{b}=\frac{q^{2}n_{2}}{2\epsilon_{s}}\left( L-W\right)^{2}+qU+\chi_{s}.
\end{equation}

Using this and the definition of the ideality factor (eq. (~\ref{def-if})) we arrive at the following expression for 
the bias dependent ideality factor in non-ideal metal-semiconductor systems:
\begin{equation}
n(U)=\frac{1}{\left(1-\frac{L}{W(U)}\right)\left(1+\frac{L d\sigma (U)}{\epsilon_{s} dU}\right)}.
\label{if}
\end{equation}

This result limits to the well known result of Card and Rhoderick~\cite{card} when $L<<W$. If we introduce depletion and
interlayer geometric capacitances $C_{d}=\epsilon_{s}/W$, $C_{i}=\epsilon_{s}/L$, we obtain:
\begin{equation}
n(U)=\frac{1+C_{d}/C_{i}}{1+\frac{d\sigma}{C_{i}dU}}.
\end{equation}

Card and Rhoderick~\cite{card} proposed the division of the interface states into two groups: one 
in equilibrium with the metal, and the other in equilibrium with the semiconductor. 
Gomila and Rubi~\cite{gomila2,gomila} have considered the
problem of the bias dependence of ideality factor in case of nonequilibrium, and using a kinetic model derived appropriate bias 
dependence of these partial densities of states.




The detailed electronic structure of the non-ideal metal-semiconductor contact must be complex involving 
possible defect, metal and disorder induced densities of states. The charging mechanisms at the interface are 
therefore quite complicated. For instance, Wu, Yang, and Evans~\cite{wu2} described the observed excess 
and negative capacitance in NiSi$_{2}$/n-Si(100) Schottky diode 
using the modified occupation function of the interface states. They introduced modification causing the 
decreasing of the interface charge with the forward bias on account of the decreasing of the occupancy of the 
interface states below the metal Fermi energy. 

Here we use a much simpler model for the complex electronic structure which however   
gives suprisingly good description of the functional bias dependence 
of the excess charge. The two defect levels are considered as two
sharp electronic levels $\epsilon_{1}$ and $\epsilon_{2}$ interacting with the partial densities of states in 
equilibrium with the metal and semiconductor. 
This mechanism is in some respect similar to the description of the interaction of 
the metal clusters on the semiconductor surface\cite{doniach}. The important 
difference is that in our case we consider a single defect coupled 
to the interface states, so that the transferred charge cannot be screened out
as it is the case in the metal cluster~\cite{doniach}. 

The random distribution of the metal atoms in the disordered interlayer 
causes the fluctuations of the potential at leads to 
broadening of the sharp defect levels. In the high density limit the
distribution function of the potential $V(\vec{r})$ can be 
approximated by the gaussian:
\begin{equation}
P(V)=\frac{1}{\sqrt{\pi}U_{0}}\exp (-V^{2}/(U_{0})^2),
\end{equation}
where the parameter $U_{0}$ from the charge model function 
was immediately recognized as the 
width of the distribution function. 
The disordered density of states in semiclassical 
approximation yields\cite{kane,mieghem}:
\begin{equation}
\rho_{1}(\epsilon)=
\frac{1}{\sqrt{\pi}U_{0}}\left( \exp (-(\epsilon-\epsilon_{1})/U_{0})^{2})\right).
\end{equation}
\begin{equation}
\rho_{2}(\epsilon)=
\frac{1}{\sqrt{\pi}U_{0}}\left( \exp (-(\epsilon-\epsilon_{2})/U_{0})^{2})\right).
\end{equation}
If the impurity density is $n$ then we obtain for the excess charge 
density using the low temperature limit $(T\to 0)$ for simplicity:
\begin{equation}
\sigma (\epsilon_{f})=\frac{q n^{2/3}}{2}\left( {\rm erf} [(\epsilon_{f}-(\epsilon_{d}-\Delta\epsilon))/U_{0}]
+{\rm erf} [(\epsilon_{d}+\Delta\epsilon-\epsilon_{f})/U_{0}]\right),
\end{equation}
where $\epsilon_{f}$ is the semiconductor Fermi energy, $n$ is the defect concentration, $\epsilon_{d}$ is the
mean defect level, and $2\Delta\epsilon$ is the separation between the two defect levels. If the shift of
the defect levels from the mean value is small compared to the disorder parameter $U_{0}$, i.e. 
$\Delta\epsilon /U_{0}<<1$, then we have approximately:  
\begin{equation}
\sigma (\epsilon_{f})\approx \frac{2q n^{2/3}\Delta\epsilon}{\sqrt{\pi}U_{0}}
\exp [-(\epsilon_{f}/U_{0}-\epsilon_{d}/U_{0})^{2}].
\label{thmodel}
\end{equation}
The difference between the Fermi energy in the metal and the Fermi energy in the semiconductor far from 
interface equals the applied bias:
\begin{equation}
\epsilon_{f}=\epsilon_{fm}+qU.
\end{equation}
In equilibrium, the Fermi energy at the interface is found to be pinned near the midgap of the semiconductor.
In one dimension it tends to fall near the branch point in the complex band structure. In three dimensions 
the pinning point is found to be given by the gap center of the minimum indirect gap, $\epsilon_{g}$, 
corrected with spin-orbit
splitting~\cite{tersoff}, and including the shift from the defined center depending on the metal $\delta_{m}$:
\begin{equation}
\epsilon_{fm}=\frac{1}{2}\epsilon_{g}+\delta_{m}.
\end{equation}
Using (4), and (5) in the expression for the interface charge density (3), one can define an effective gap center,
$\epsilon_{0,eff}$, in the presence of the disorder and defects in the semiconductor: 
\begin{equation}
\epsilon_{0,eff}=KU_{0}=\frac{1}{2}\epsilon_{g}+\delta_{m}-\epsilon_{d}=\frac{1}{2}\epsilon_{g}-\delta\epsilon_{m,d}.
\end{equation}
This shows that the parameter $K$ of the model charge function (1) identifies the position of the effective gap center at 
non-ideal metal-semiconductor interface. The Schottky barrier height is in most models describing the 
Fermi level pinning~\cite{bardeen,hasegawa,tung00} given as an interpolation between the Schottky-Mott and
Bardeen limit: 
\begin{equation}
\varphi_{b,p}=c(I_{s}-\phi_{m})+(1-c)\epsilon_{0,eff}.
\end{equation}
Here, $I_{s}=E_{g}+\chi_{s}$ is the semiconductor ionization potential, and $c$ is the pinning parameter which 
depends on the details of the pinning mechanism. As the pinning parameter changes its value from 1 to 0, the 
Schottky barrier height goes from totally unpinned to fully pinned value. 

\section{Results and discussion}

Using the experimentaly observed 
values for the Schottky barrier height on p-type Si and 
the effective gap center deduced from the analyzed capacitance-voltage
characteristics~\cite{cvikl02} one can obtain the pinning parameter for different metal-semiconductor systems
that were analyzed~\cite{cvikl02}. 
If the pinning mechanisms depends on the continuum of the localized density of states in the band gap, then 
the pinning parameter of the following form is usually obtained~\cite{rhoderick} for nearly ideal systems:
\begin{equation}
c=\frac{1}{1+\frac{q^{2}D(\epsilon_{0,eff})\xi}{\epsilon_{s}}},
\end{equation}
where $D(\epsilon_{0,eff})$ is the (two dimensional) density of states in the band gap at the effective gap
center, $\xi$ is the localization length of the states, and $\epsilon_{s}$ is the dielectric constant
of the semiconductor. 

When the disordered interlayer of width $L$ is introduced between the metal and the bulk semiconductor, 
possibly extending over several atomic layer, the pinning parameter is given by~\cite{hasegawa}:
\begin{equation}
c={\rm sech}\left(\sqrt{q^2 L^2 \rho (\epsilon_{0,eff})/\epsilon_{s}}\right),
\end{equation}
where $\rho (\epsilon_{0,eff})$ is the density of disorder induced gap states. 

The energy interval determining the properties of the non-ideal metal-semiconductor
junction is estimated to be of the order of $\Delta\epsilon$. 
The density
of the defect levels interacting with the continuum of localized states is then approximately given by: 
\begin{equation}
D(\epsilon_{0,eff})\Delta\epsilon\approx n^{2/3}.
\end{equation}
Using (8) and (9) in eq. (3) gives the final expression for the interface charge density in nearly ideal systems:
\begin{equation}
\sigma (U)=\frac{2 q^{3}n^{4/3}\xi}{(1/c-1)\sqrt{\pi}\epsilon_{s}U_{0}}
\exp [-(\epsilon_{0,eff}/U_{0}+qU/U_{0})^{2}],
\end{equation}
and the corresponding expression for the non-ideal systems:
\begin{equation}
\sigma (U)=\frac{2 q^{3}n^{4/3}L}{({\rm arcsech(c)})^{2}\sqrt{\pi}\epsilon_{s}U_{0}}
\exp [-(\epsilon_{0,eff}/U_{0}+qU/U_{0})^{2}]
\end{equation}
The expression for the interface charge density (eq. 10) as derived from the presented model shows that the crucial
microscopic parameters detemining the interface charge density in non-ideal metal-semiconductor contacts are:
the disorder parameter $U_{0}$, density of defects $n$, Schottky barrier pinning parameter $c$ (or
the density of states interacting with the defect level and localization length), and the effective gap center. 
The parameters $b$ (the maximum of the interface charge density) and $U_{0}$ obtained from 
the analysis of the capacitance-voltage characteristics of non-ideal metal-semiconductor contacts~\cite{cvikl02}
allows one to compute the number of defects per unit area, the density of corresponding continuum of states 
and the shift of the defect level interacting with
the continuum of localized states. The results of the calculation are presented in tables 1 and 2. 
The calculated effective gap center for the series
of metals on p-type Si (Ag, Pb, Al, Ti, Mo) reasonably well 
corresponds to the calculated avarage branch point, $\epsilon_{0}=0,36$ eV, 
in the Si complex band structure~\cite{tersoff}, the maximum deviation being 0,1 eV. 

We have also considered the negative and excess capacitance observed by Wu, Yang, and Evans~\cite{wu2} in their samples. 
The excess capacitance is in the case $L<<W$ according to
eq.~\ref{cap} simply equal to derivative of the
interface charge with respect to applied bias: $C_{excess}\approx -d\sigma /dU$. Using generalized model for the 
bias dependence of the interface charge eq.(~\ref{thmodel}):
\begin{equation}
\sigma (U)=q n_{1}{\rm erf}\left( \epsilon_{0}+\Delta\epsilon-qU)/U_{0,1}\right)+
q n_{2}{\rm erf}\left(qU-\epsilon_{0}+\Delta\epsilon )/U_{0,2}\right),
\label{cf2}
\end{equation}
we are able also to describe the results of Wu, Yang and Evans~\cite{wu2} within the framework of
our model as it is shown in figure 2. The parameters used in calculation were: $\epsilon_{0}=0,5$ eV, $\Delta\epsilon =-0,1$ eV,
$U_{0,1}=0,065$ eV, $U_{0,2}=0,17$ eV, $n_{1}=3,8\times 10^{14}$ cm$^{-2}$ and $n_{2}=3,1\times 10^{14}$ cm$^{-2}$.

We now apply the derived results for the bias dependent ideality factor given by eq. (~\ref{if}) to two sets of published data 
measured in quite different non-ideal metal-semiconductor systems. 

Maeda et al.~\cite{maeda} have observed bias dependent ideality factor in Au/n-GaAs metal-semiconductor system.
The bias dependence of the ideality factor exhibited a pronounce peak at certain applied bias. We have used 
eqs. (~\ref{cf}) and (~\ref{if}) and calculated the measured bias dependence of the ideality factor using the
following set of parameters: $b=4,6\times 10^{-3}$ As/m$^{2}$, $U_{0}=0,33$ eV, $K=0,042$, $L=2$ nm. Figure 3
shows the measured set of data and the calculated bias dependence of the ideality factor. In tables 1 and 2 
we present the computed density of defects, the shift of defects energy levels and the density of states. 

In figure 4 we present the experimentally observed bias dependence of ideality factor as measured in Au/n-InP 
Schottky structure obtained from~\cite{ahaitouf} (empty dots). 
In the applied voltage interval for the presented set of data~\cite{ahaitouf} we can write for the derivative of the 
charge density:
\begin{equation}
-\frac{Ld\sigma}{\epsilon_{s}dU}=c_{1}\exp\left(-(-qU/c_{2}+c_{3}/c_{2})^{2}\right)+c_{5}+c_{4} U,
\end{equation}
with the following values of parameters: $c_{1}=0,5$, $c_{2}=U_{0}=0,18$ eV, 
$c_{3}=\epsilon_{d}=0,11$ eV, $c_{4}=0,25$ 1/V, $c_{5}=0,19$.
The width of the interlayer was $L=15$ nm~\cite{ahaitouf}.
From these values we 
calculate the density of the defects $N_{d}=2,8\times 10^{11}$ cm$^{-2}$ and the barrier height for 
Au/InP contact: $\varphi_{b,n}=0,76$ eV which agrees well with the results obtained from 
analysis of I-V curves in ref.~\cite{ahaitouf}. 

\section{Conclusions}
In this work we have shown that the non-ideality in metal-semiconductor contacts which is observed either in 
current-voltage as bias dependent ideality factor or in capacitance voltage characteristics as 
excess capacitance can be described using a single model for the bias dependent interface charge. 

The expression for the interface charge density is derived which showed that the crucial
microscopic parameters detemining the interface charge density in non-ideal metal-semiconductor contacts are:
the disorder parameter $U_{0}$, density of defects $n$, Schottky barrier pinning parameter $c$ (or
the density of states interacting with the defect level and localization length), and the effective gap center $\epsilon_{0,eff}$. 
The parameters $b$ (the maximum of the interface charge density) and $U_{0}$ obtained from 
the analysis of the capacitance-voltage characteristics of non-ideal metal-semiconductor contacts~\cite{cvikl02}
allows one to compute the number of defects per unit area and the shift of the defect level interacting with
the continuum of localized states. 
The basic assumption was to condsider 
the random distribution of the metal atoms in the disordered layer 
causes the fluctuations of the potential in the layer resulting in 
broadening of the sharp defect levels. In the high density limit where the
distribution function of the potential can be approximated by the gaussian. 

The bias dependent charge residing at the metal-semiconductor interface is connected to the bias dependence of the
ideality factor of the current-voltage characteristic. The comparison of the results computed from the 
theoretical model with the published experimental data enabled us to extract several microscopic properties of 
various metal-semiconductor contacts.

\newpage


\begin{figure}[h]
\begin{center}
\includegraphics[width=12cm,keepaspectratio]{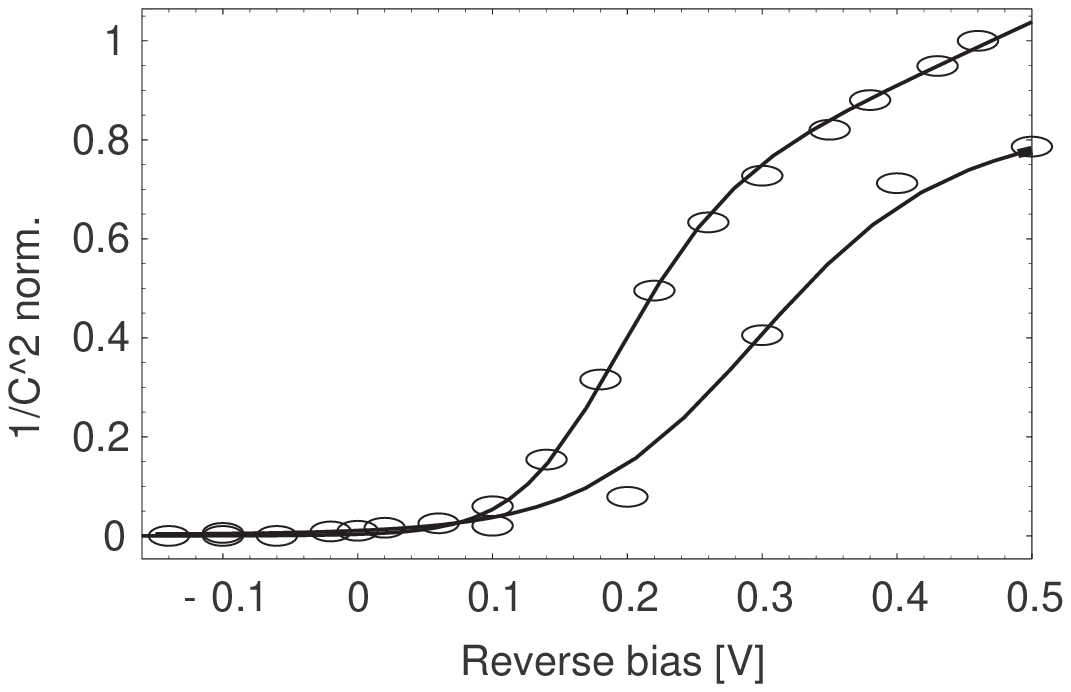}
\caption{Low frequency capacitance of e-beam deposited, Al/n-Si
as a function of the external reverse bias~\cite{tseng1,tseng2}, and
low frequency capacitance of ICB deposited
Pb/p-Si(100) as a function of the external reverse bias
are compared
to the calculated values~\cite{cvikl02}.}
\end{center}
\end{figure}

\newpage

\begin{figure}[h]
\begin{center}
\includegraphics[width=12cm,keepaspectratio]{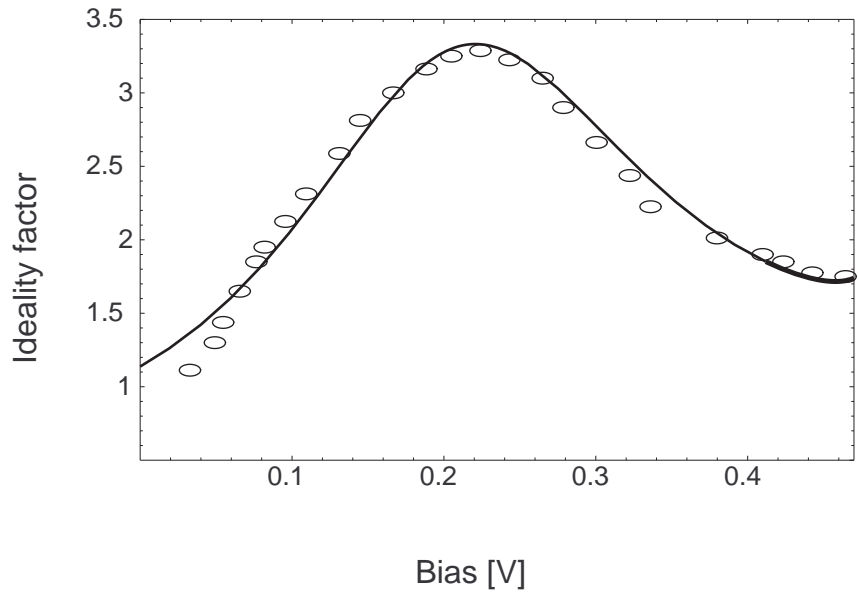}
\caption{Values of ideality factor (empty dots) in Au/n-GaAs(100)
as obtained from experiment~\cite{maeda}, and 
calculated bias dependence of ideality factor (solid line).}
\end{center}
\end{figure}

\newpage

\begin{figure}[h]
\begin{center}
\includegraphics[width=12cm,keepaspectratio]{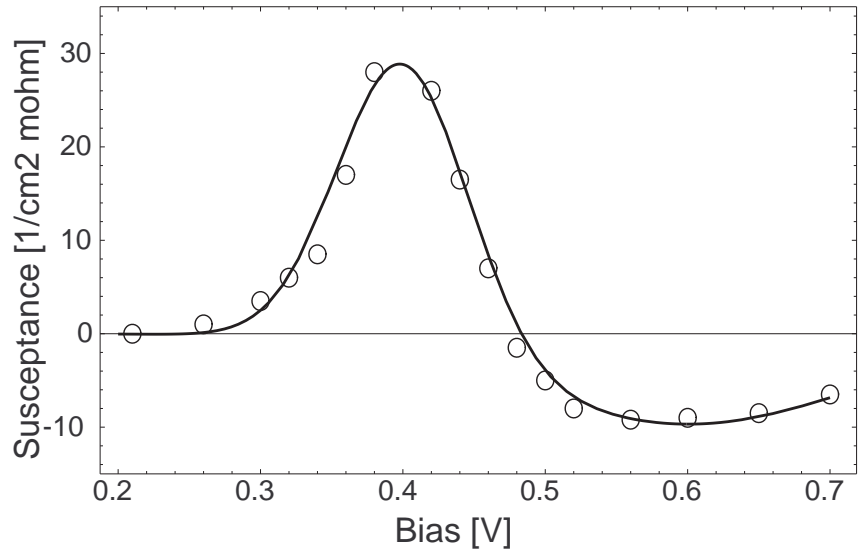}
\caption{Measured low frequency susceptance as a function of applied bias (empty dots) of e-beam deposited
NiSi$_{2}$/n-Si(111)~\cite{wu1,wu2} and calculated susceptance (solid line) using the two-level model for the bias
dependent interface charge.}
\end{center}
\end{figure}

\newpage

\begin{figure}[h]
\begin{center}
\includegraphics[width=12cm,keepaspectratio]{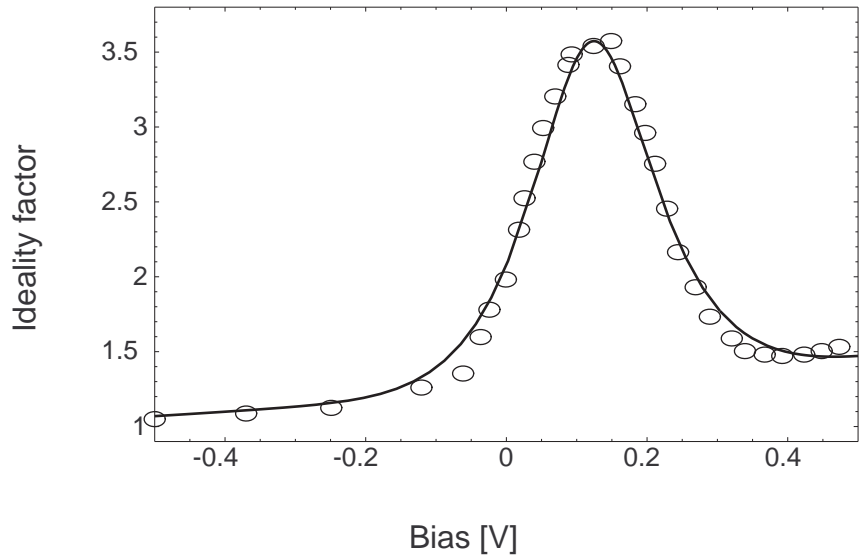}
\caption{Values of ideality factor (empty dots) in Au/n-InP(100)
as obtained from experiment~\cite{ahaitouf}, and 
calculated (solid line) bias dependence of ideality factor.}
\end{center}
\end{figure}

\newpage

\begin{table}[h]
\begin{center}
\begin{tabular}{ccccccc}\hline \hline
metal & $\varphi_{b,SM} [eV] $ & $\epsilon_{0,eff} [eV] $ & $\varphi_{b,p} [eV] $ 
& c & D [eV$^{-1}$cm$^{-2}$]$\times 10^{14}$ & $\rho$ [eV$^{-1}$cm$^{-3}$]$\times 10^{19}$\\ \hline
Ag & 0,75 & 0,35 & 0,6  & 0,63 & & 1,1\\
Pb & 1,13 & 0,39 & 0,73 & 0,46 & & 2,1\\
Al & 0,97 & 0,43 & 0,54 & 0,2 & 10,6 & \\
Ti & 1,07  & 0,3  & 0,53 & 0,29 & 6,5 & \\
Mo & 0,96 & 0,26 & 0,48 & 0,31 & 5,9 &\\
Ag(1 kV) & 0,75 & 0,26 & 0,54 & 0,57 & & 3,6\\
Au/GaAs & 0,79 & 0,01 & 0,56 & 0,70 & 1,1 &\\ \hline \hline
\end{tabular}
\end{center}
\caption{\sl Values of experimentaly determined effective gap center, barrier height and calculated values of
the pinning parameter and the density of states. 
The values for the second column were taken from~\cite{rhoderick,sze}, and for the
third and fourth column from~\cite{cvikl02}. In the calculation the localization length was taken to be
0,25 nm for the nearly ideal metal-semiconductor contacts. In non-ideal cases the width of the interlayer was
used.} 
\end{table}

\newpage

\begin{table}[h]
\begin{center}
\begin{tabular}{ccccc}\hline \hline
metal & $b\times 10^{-3}$ [As/m$^{2}$] & $U_{0}$ [eV] & $N_{d}\times 10^{12}$ [cm$^{-2}$] & $\Delta\epsilon$ [meV] \\ \hline
Ag & 1,4 & 0,5 & 7,8  & 28 \\
Pb & 0,4 & 0,3 & 4,5 & 8,2 \\
Al & 0,37 & 0,16 & 5,9 & 5,6 \\
Ti & 0,2  & 0,12  & 2,9 & 4,4 \\
Mo & 0,41 & 0,16 & 4,6 & 7,8 \\
Ag(1 kV) & 3,2 & 0,2 & 8,4 & 23,3 \\ 
Au/GaAs & 4,6 & 0,33 & 5,5 & 50 \\ \hline \hline
\end{tabular}
\end{center}
\caption{\sl Values of experimentaly determined maximum of the interface charge density $b$, 
disorder parameter, and calculated
defect density and energy level shift.} 
\end{table}

\end{document}